\documentclass[letter]{aa} 

\usepackage{graphicx}
\usepackage{layouts}
\usepackage{txfonts}
\usepackage{natbib}

\begin{document}

   \title{Tuning in on Cepheids: Radial velocity amplitude modulations 
   \thanks{Based on observations collected at ESO La Silla Observatory using
   the {\it Coralie} spectrograph mounted to the Swiss 1.2m Euler telescope.
   The derived radial velocities are available in electronic form at the CDS via
   anonymous ftp to cdsarc.u-strasbg.fr (130.79.128.5) or
   at \texttt{http://cdsweb.u-strasbg.fr/cgi-bin/qcat?J/A+A/}}}

   \subtitle{A source of systematic uncertainty for Baade-Wesselink distances}

   \author{Richard I. Anderson}

   \institute{Observatoire de Gen\`eve, Universit\'e de Gen\`eve, 51 Ch. des
   Maillettes, 1290 Sauverny, Switzerland\\
    \email{richard.anderson@unige.ch} }

   \date{Received 20 March 2014; accepted 3 June 2014}

 
  \abstract
   {Classical Cepheids are crucial calibrators of the extragalactic distance
   scale. The Baade-Wesselink technique can be used to calibrate Cepheid
   distances using Cepheids in the Galaxy and the Magellanic Clouds.}
   {I report the discovery of modulations in radial velocity (RV) curves of
   four Galactic classical Cepheids and investigate their impact as a
   systematic uncertainty for Baade-Wesselink distances.}
   {Highly precise Doppler measurements were obtained using the {\it Coralie}
   high-resolution spectrograph since 2011.
   Particular care was taken to sample all phase points in order to very
   accurately trace the RV curve during multiple epochs and to search for
   differences in linear radius variations derived from observations obtained at
   different epochs. Different timescales are sampled, ranging from
   cycle-to-cycle to months and years.}
   {The unprecedented combination of excellent phase coverage 
   obtained during multiple epochs  and high precision enabled the
   discovery of significant modulation in the RV curves of the short-period {\it s-}Cepheids QZ\,Normae and V335\,Puppis, as well as the
   long-period fundamental mode Cepheids $\ell$\,Carinae and RS\,Puppis.
   The modulations manifest as shape and amplitude variations that vary smoothly
   on timescales of years for short-period Cepheids and from one pulsation
   cycle to the next in the long-period Cepheids.
   The order of magnitude of the effect ranges from several hundred m\,s$^{-1}$
   to a few km\,s$^{-1}$. The resulting difference among linear radius
   variations derived using data from different epochs can lead to systematic
   errors of up to $15\%$ for Baade-Wesselink-type distances, if the employed
   angular and linear radius variations are not determined contemporaneously. }
   {The different nature of the Cepheids exhibiting modulation in their  RV
   curves suggests that this phenomenon is common. The observational baseline is not
   yet sufficient to conclude whether these modulations are periodic. To ensure the accuracy of Baade-Wesselink distances, angular and linear
   radius variations should always be determined contemporaneously.}


   \maketitle


\section{Introduction}
Classical Cepheids are crucial distance tracers over the range from several
hundred parsecs up to one hundred Megaparsec. Thanks to this, Cepheids and
type\,Ia supernovae together allow a one-step calibration of the Hubble
constant, $H_0$ \citep[e.g.,][]{2011ApJ...730..119R,2012ApJ...758...24F}. Such
extragalactic distances are estimated using the Cepheid period-luminosity
relation (PLR), which was originally discovered by Henrietta Leavitt
\citep{1908AnHar..60...87L,1912HarCi.173....1L}. The calibration of this
relationship is very important for astronomy, the distance scale,
and cosmology, since about half of the methods promising percent precision on a
calibration of $H_0$ are Cepheid-related \citep{2013IAUS..289....3F}.

Accurate Cepheid distances are required to achieve an accurate calibration of
the PLR. This endeavor has a rich history and much literature can be found on
the subject \citep[see][]{1999PASP..111..775F}. 
For the Galactic PLR calibration, 
there are essentially three methods that can provide good distance estimates.
The gold standard among these are trigonometric parallaxes; see notably
\citet{1997MNRAS.286L...1F} and \citet{2007MNRAS.379..723V}, who used the
parallaxes measured by the {\it Hipparcos} space mission.
\citet{2002AJ....124.1695B,2007AJ....133.1810B} employed the {\it Hubble Space
Telescope} to measure highly precise parallaxes of ten Cepheids.
Within the next two to eight years, the recently launched space mission {\it Gaia}
will provide Cepheid parallaxes of unprecedented accuracy (tens of $\mu$arcsec),
and the program by \citet{2014ApJ...785..161R} holds great
promise to determine parallaxes of Cepheids within $5$\,kpc with similar accuracy. 

Another important means of calibrating the PLR is using Cepheids
associated with open clusters; for example, see \citet{2002AJ....124.2931T}, \citet{2010Ap&SS.326..219T}, and \citet{2013MNRAS.434.2238A}. In this
case, host cluster distances provide independent distance estimates for member 
Cepheids.

Finally, Cepheid distances can be determined by exploiting the pulsations via
different variants of the Baade-Wesselink (BW) technique
\citep{1926AN....228..359B,1940ZA.....19..289B,1946BAN....10...91W}. In this
way, precise distances to many (>100) Cepheids in the Galaxy and the Magellanic
Clouds have been determined
\citep{1993ApJ...418..135G,1998ApJ...496...17G,2004A&A...415..531S,2011A&A...534A..95S,2007A&A...476...73F,2008A&A...488...25G,2013A&A...550A..70G}.
Recently, the infrared version of the surface brightness technique
\citep{1975MNRAS.172..455T,1976MNRAS.174..489B} has become the most successful
in terms of precision, since it is calibrated using interferometric measurements of
red giant stars and supergiants \citep{1997A&A...320..799F}, as well as classical
Cepheids \citep{2004A&A...428..587K}. Using the VLTI,
\citet{2004A&A...423..327K} were able to interferometrically measure angular
variations due to pulsation and achieved a PLR calibration based on eight
Galactic Cepheids. %

BW distances are determined by measuring angular 
(e.g., via interferometry, or optical \& near-IR photometry) 
and linear (via Doppler measurements) radius variations.
Following \citet{2001A&A...367..876K} the distance is computed as follows:
\begin{equation}
d\,\rm{[pc]} = 9.305 \cdot \frac{2 \Delta R}{\Delta \Theta}\, ,
\label{eq:IRSBdist}
\end{equation}
where $\Delta R$ $[R_\odot]$ denotes the linear radius variation, and $\Delta
\Theta$ [mas] is the variation in angular diameter.

It is well known that both $\Delta \Theta$ and $\Delta R$ should be
measured during the same pulsation cycle. In practice, however, this is not
always possible owing to telescope time restrictions, especially when working with
interferometry or statistical data sets that should yield the most precise PLR
calibrations. Furthermore, Cepheid pulsations are usually considered to be
rather regular (apart from well-known period changes and a few peculiar cases,
such as HR\,7308 \citep{1982A&A...109..258B} or Polaris, cf. \citealt{2009AIPC.1170...59T} and
references therein).
For instance, \citet{1997MNRAS.292..662T} found no indication of cycle-to-cycle
variations in radial velocity (RV) curves at the level of 600\,m\,s$^{-1}$ in
$\ell$\,Car, and \citet{2004ApJ...603..285M} conclude that such an effect is
negligible compared to other sources of systematic uncertainty. 
The present work, however, demonstrates that significant RV modulations can be
found in some Cepheids not usually considered to be peculiar.

Modulation can lead to systematic errors in Baade-Wesselink
distance estimates if $\Delta \Theta$ and $\Delta R$ in
Eq.\,\ref{eq:IRSBdist} are determined using data from different pulsation
cycles.
Consider that $\Delta R$ is computed using a projection factor, $p$,
as
\begin{equation}
\Delta R = p \cdot \int{v_r dt}\,.
\label{eq:DeltaR}
\end{equation}
Assuming $p$ does not vary between pulsation cycles, modulation will
result in different $\Delta R$ determined from data observed at
different epochs.
We furthermore assume that $\Delta \Theta$ is also subject to modulations
(i.e., they relate to the photospheric radius) and is determined during epoch 1,
then we can quantify
the relative distance error resulting from using $\Delta R$ measured during
epoch 2 instead of epoch 1 as 
\begin{equation}
err(d) \equiv \frac{d_1 - d_2}{d_1} = \frac{\Delta R_1 - \Delta R_2}{\Delta R_1}
= \frac{\int_1{v_r} - \int_2{v_r}}{\int_1{v_r}}\,.
\label{eq:errD}
\end{equation}
This error is independent of the value of $p$.
In the following, the integral is referred to by its equivalent, $\Delta R/p$.

In this \emph{letter}, I report on newly-discovered RV modulations and
estimate the systematic error that results from employing non-contemporaneous 
$\Delta R$ and $\Delta \Theta$ according to the above reasoning. Following an
overview of the observational setup in Sect.\,\ref{sec:obs}, RV modulations
and systematic BW distance error estimates are presented in
Sect.\,\ref{sec:res}. 
Section\,\ref{sec:disc} discusses possible origins of this phenomenon, and 
Sect.\,\ref{sec:conclu} provides brief conclusions. Supporting
figures are provided in the online appendix.
   
\section{Observations}\label{sec:obs}
The radial velocities presented here were determined from observations
taken between April 2011 and  February 2014 using the fiber-fed
high-resolution (R $\sim$ 60\,000) spectrograph {\it Coralie} at the 
Swiss 1.2m Euler telescope located at ESO La Silla Observatory in Chile.
{\it Coralie} is described in \citet{2001Msngr.105....1Q}. 
\citet{2010A&A...511A..45S} provide a description of instrumental updates
made in 2007.
An efficient reduction pipeline is available for {\it Coralie}. The reduction
follows standard procedure and performs pre- and overscan bias correction,
flatfielding using Halogen lamps, and background
modelization, as well as cosmic removal. ThAr lamps are used for the
wavelength calibration.

RVs were determined via cross-correlation 
\citep{1996A&AS..119..373B,2002A&A...388..632P} using a
numerical mask designed for solar-like stars (optimized for spectral type G2).
The instrument is tried and tested, and is renowned for its stability and very
high precision of $\sim 3$\,m\,s$^{-1}$
\citep{2003ASPC..294...39P,2010A&A...511A..45S}. 

All data used here are made
publicly available at the
CDS\footnote{\url{http://cds.u-strasbg.fr/}}

\section{Results}\label{sec:res}

\begin{table*}
\centering
\begin{tabular}{@{}ll|rrrrrrrrrr@{}}
& & S/N &
 N$_1$ & N$_2$ & $\Delta$t & BJD$_1$  & BJD$_2$ & $\Delta A_{v_{r}}$ & $\Delta
 R_{1}/p$ &  $\Delta R_{2}/p$ & err(d) \\
Cepheid & HD & & & & [yr] & [d] & [d] & [km\,s$^{-1}$] & [$R_\odot$]
& [$R_\odot$] & [\%] \\
\hline
QZ Nor & 144972 & 29 &  18 &  16 & 2.1 & 5660.3 & 6430.3 & $-$1.26 
$\pm$ 0.05 & 1.10 $\pm$ 0.01 & 1.26  $\pm$  0.01 & $-$14.5  $\pm$  0.6  \\
V335 Pup & 65227 & 36 &  11 &  30 & 2.8 & 5659.6 & 6660.8 & 1.12 
$\pm$ 0.01 & 1.06 $\pm$  0.01 & 0.95  $\pm$  0.01 & 10.3 $\pm$  0.2\\
l Car & 84810 & 257 & 110 & 151 & c2c & 6660.7 & 6701.7 & $-$0.51 
$\pm$ 0.06 & 22.42 $\pm$  0.01 & 23.62  $\pm$  0.01 & $-$5.1  $\pm$  0.1 \\
 & & 263 & 37 &  86 & 1.7 & 6089.5 & 6659.7 & 1.35  $\pm$  0.31 &
 23.95 $\pm$  0.02 & 22.42  $\pm$  0.01 & 6.8  $\pm$  0.1 \\
RS Pup & 68860 & 133 & 124 & 170 & c2c & 6660.8 & 6703.6 & 1.32 
$\pm$ 0.11 & 35.38  $\pm$  0.01 & 33.93  $\pm$  0.01 & 4.3  $\pm$  0.1  \\
& & 125 & 29 & 131 & 1.1 & 6327.6 & 6661.6 & $-$3.03  $\pm$  0.08 &
32.79 $\pm$  0.12 & 35.42  $\pm$  0.01 & $-$7.4  $\pm$  0.4 \\
\hline
\end{tabular}
\caption{
Estimating the possible impact of modulations on BW distances 
using RV data from different epochs.
The columns list: median signal-to-noise ratio (S/N) of the spectra at
$5700 \AA$; number of observations per epoch (N$_{1,2}$);
baseline ($\Delta$t) between the epochs considered (c2c indicates
two consecutive pulsation cycles); median
Barycentric Julian dates, less $2\,450\,000$ (BJD$_{1,2}$).
$\Delta A_{v_r}$ is the change in peak-to-peak RV amplitude due to
modulation.
Columns $\Delta R_{1,2}/p$ list the per-epoch RV curve integrals, cf.
Eq.\,\ref{eq:DeltaR}. The last column (err(d)) quantifies the 
systematic uncertainty on BW distances due to modulation if non-contemporaneous $\Delta R$ and $\Delta
\Theta$ are used to determine the  distance, cf. Eq.\,\ref{eq:errD}.}
\label{tab:obs}
\end{table*}

\begin{figure*}
\includegraphics{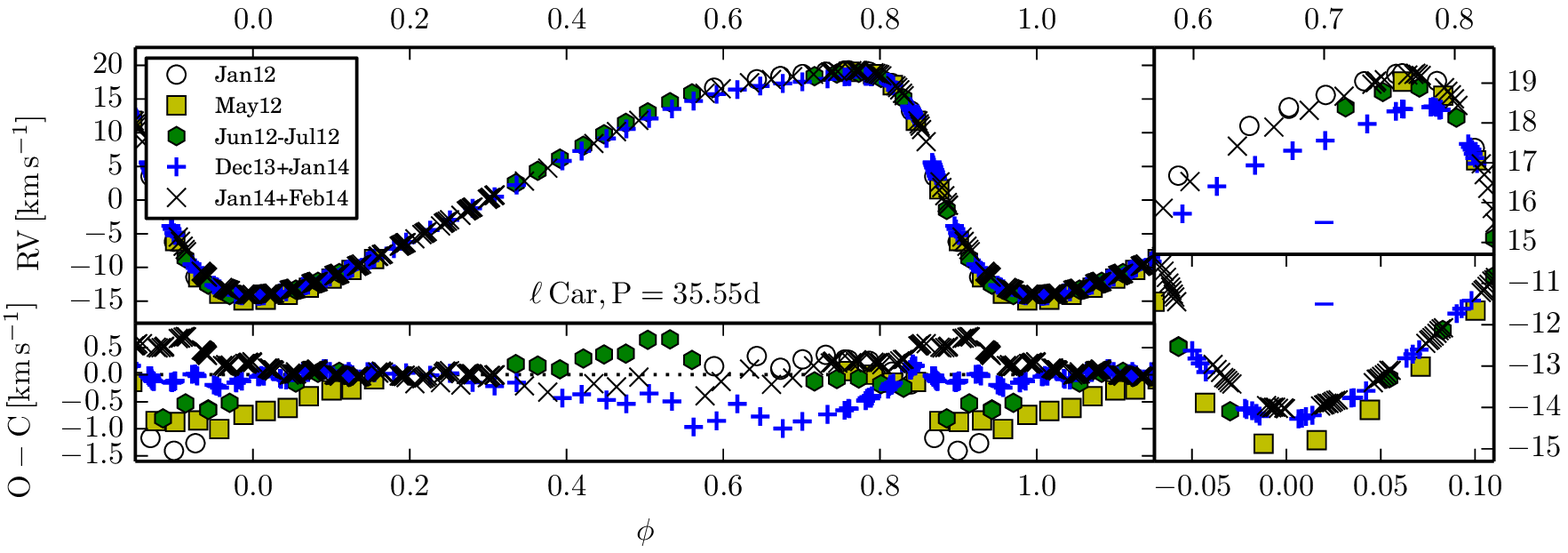}
\caption{New {\it Coralie} RV data for $\ell$~Car. The large
panel shows the phased RV curve, distinguishing data from different
epochs by symbol style and color. The
two right-hand panels provide close-ups around maximum (upper) and minimum RV
(lower), and show the median RV uncertainty as an errorbar (too small to
discern in this case).
The bottom panel shows residuals around an
``average'' Fourier series model fit to the combined data set
\citep[fit procedure described in][]{2013MNRAS.434.2238A}. Figures for the
other Cepheids are provided in the online version.}
\label{fig:lCar}
\end{figure*}

Using the new {\it Coralie} data, modulations in shape and amplitude
of RV curves are discovered in four Cepheids of rather different natures
(pulsation mode and period, mass and radius). 
Figure\,\ref{fig:lCar} shows the modulations for one of these,
$\ell$~Car. Figures for the other three targets are provided in the online
appendix. Interestingly, the short-period Cepheids exhibit smooth long-term
modulations, whereas the long-period Cepheids exhibit variations between subsequent pulsation
cycles. More details on this and other features of the modulations will be
presented in a forthcoming publication.

To estimate the impact of RV modulation on BW distances, data from two
epochs are used following Eq.\,\ref{eq:errD}.
Table\,\ref{tab:obs} summarizes the observational data and results employed in
this estimation, and lists the peak-to-peak RV amplitude variations between
epochs ($\Delta A_{v_r}$), the integrals of the RV curves ($\Delta R/p$), and
the relative distance error as defined in Eq.\,\ref{eq:errD}.
To compute the integrals, the per-epoch data were represented by splines
for QZ~Nor, $\ell$~Car, and RS~Pup. A second-order Fourier series was more
appropriate for V335~Pup. The central values and uncertainties for
$\Delta A_{v_r}$, $\Delta R/p$, and $\rm{err(d)}$ were then determined using a
classical Monte Carlo method, in which the analysis was repeated 10\,000
times using randomly offset datapoints (offsets drawn from a normal distribution
with variance equal to the squared measurement uncertainties).

\vspace{1mm}{\bf \object{QZ Normae}}\hspace{0.5cm}The {\it s-}Cepheid QZ\,Normae
resides in the open cluster NGC\,6067 \citep[see][and references
therein]{2013MNRAS.434.2238A}, making it an important calibrator for the
Galactic PLR.
Interestingly, the modulations are
asymmetric around the mean and are significantly larger during
contraction than during expansion.
The smoothly and steadily increasing amplitude is traced closely over an
observational baseline of nearly three years. Two well-traced epochs lying
$2.1$ years apart yield significantly different values for $\Delta R$
(increasing with time), leading to a possible systematic distance error of
nearly $15\%$.

\vspace{1mm}{\bf \object{V335 Puppis}}\hspace{0.5cm}
The modulations in this {\it s-}Cepheid  
smoothly decreased RV amplitude over a period of nearly three years, before a
reversal became apparent in February 2014. This contrasts with 
the modulation in QZ\,Nor, which thus far has not shown a reversal. Using the
two most extreme (in terms of amplitude) pulsation cycles yields $\Delta R$
values that differ by $11\%$.
%

\vspace{1mm}$\mathbf{\ell}$\,{\bf Carinae}\hspace{0.5cm}
The long-period Cepheid $\ell$\,Carinae (\object{HIP 47854}) is particularly
important for the distance scale. Firstly, the frequently adopted Galactic
PLR calibration by \citet{2007AJ....133.1810B} relies heavily on this
star, since it is the only long-period (P>10d) Cepheid in their sample.
Secondly, $\ell$\,Car's pulsations can be resolved by the VLTI
\citep{2004A&A...416..941K}, which enables an important cross-check of its
distance via the interferometric BW method.

Significant modulations exceeding 1\,km\,s$^{-1}$ are evident in the data, 
cf. Fig.\,\ref{fig:lCar}, where 
$\ell$~Car exhibits modulations on short timescales (cycle-to-cycle), as
demonstrated by observations taken between December 2013 and February 2014.
These modulations can vary $\Delta R/p$ between subsequent cycles by up to
$5\%$. Modulations on longer timescales can be even larger, see the second
row for $\ell$~Car in Table\,\ref{tab:obs}. Therefore, modulations in
$\ell$\,Car seem to be present on all time scales, setting a
stringent constraint to observe linear and angular variations during the same
pulsation cycle.

\vspace{1mm}{\bf \object{RS Puppis}}\hspace{0.5cm}
The long-period Cepheid RS\,Pup is a particularly interesting object due to its
location in a large reflection nebula
\citep{1961PASP...73...72W,2008A&A...480..167K,2012A&A...541A..18K,2008MNRAS.387L..33F}
and erratic period changes \citep{2009AstL...35..406B} that are also
clearly seen in the present data.
As the {\it Coralie} data shows, using non-contemporaneously determined $\Delta
R$ and $\Delta \Theta$ can lead to systematic errors of up to $7\%$ for its BW
distance. Similar to $\ell$~Car, significant modulations occur even
between subsequent cycles.
This remarkable object requires a detailed discussion that is beyond the scope of this
letter.

\section{Discussion}\label{sec:disc}
For the time being, no firm conclusions can be drawn regarding the origin of
the modulations. 
Given that the modulation time scales are very different for the short (steady
over years) and long-period Cepheids (cycle-to-cycle), it seems likely
that different mechanisms are at work. A longer time baseline is
required for firmer conclusions. 

For the short-period Cepheids, it is tempting to speculate about 
the presence of a \citet{1907AN....175..325B} effect, which is known in 
RR\,Lyrae stars; see also the ``Blazhko Cepheids'' mentioned by
\citet{2008AcA....58..163S}.
It will be interesting to compare the properties of Cepheids exhibiting
modulated RV curves with Blazhko RR Lyrae stars. Another explanation could be
secular radius variations; however, V335\,Pup exhibits a reversal of the modulation, which 
may point to a recurrent phenomenon. Another possibility are non-radial
pulsations that may manifest themselves as cycle-to-cycle variations in RVs 
\citep{2003A&A...401..661K,2014A&A...561A.151N}. Strange pulsation modes 
\citep{1997A&A...326..669B,2001ApJ...555..961B}  could also provide an
explanation, since the predicted amplitudes are similar to the
observed modulations.  

One exciting possibility for the long-period Cepheids could be that their
modulations reveal the coupling between convection and pulsations. This is
plausible, given that these Cepheids are cooler and have very deep convective
envelopes. If this is the case, then $\ell$\,Car and RS\,Pup provide important
constraints for the modelization of the cool edge of the instability strip and
hydrodynamical models \citep[e.g.,][]{2013MNRAS.435.3191M}.

Another possible explanation of the cycle-to-cycle nature could be variations in
the observed stellar disk due to surface inhomogeneities (spots) moving in and
out of the field of view. 
If spots are sufficiently long-lived, a long-term (quasi-)periodicity of the
modulation might indicate such an effect.
However, short-period Cepheids have rotation periods on the order of five months
(assuming $30\,R_\odot$ and equatorial velocity $10\,\rm{km\,s^{-1}}$), which is
shorter than the observed modulation timescale (several years). Conversely,
long-period Cepheids exhibit shorter modulation timescales (cycle-to-cycle),
while their rotation periods are significantly longer (years) due to large
radii.
Since rotation has important evolutionary effects on Cepheids
\citep{2014A&A...564A.100A}, it would be particularly useful to obtain
additional information about rotation periods and velocities.

Cepheids have highly complex atmospheres and exhibit strong velocity gradients
\citep[e.g.,][]{1969MNRAS.145..377D,1993ApJ...415..323B,2006A&A...453..309N}.
Since the RVs determined here are derived from a chosen set of spectral lines,
it is possible that the RV modulations point towards long-term variations in the
time-dependent velocity structure of Cepheid atmospheres. Further investigation
in this direction is currently in progress.

It is currently unclear whether RV modulations are mirrored by modulations
in $\Delta \Theta$. To search for such signs using surface brightness methods, a
long-term photometric precision of $\sim 10\,$mmag in both V and K-band is
required. \citet{2014IAUS..301...55E} have recently reported on ``flickering'' in
space-based photometric observations of RT\,Aurigae at the level of
$20-40$\,mmag. The interferometric precision achieved for $\ell$\,Car
\citep{2004A&A...416..941K} may be sufficient to detect this modulation directly
using multi-epoch VLTI observations. A systematic search for modulations using
photometry and interferometry is thus indicated to better understand the
phenomenon and mitigate its impact on the BW technique.


\section{Conclusions}\label{sec:conclu}
This \emph{letter} presents newly discovered modulations in the radial
velocity curves of four classical Cepheids. The modulations reveal additional
complexity in the pulsation of these fundamental distance tracers. 
The four Cepheids exhibiting modulations have very different natures, two with
short and two with long periods. This fact is suggestive of a) the phenomenon
being common among Cepheids and/or b) different mechanisms being at work
(different time scales for long and short-period Cepheids).

Modulations can create significant systematic uncertainty for BW
distances, if non-contemporaneous data are employed. If RV modulations are 
mirrored by photospheric radius variations, they should be detectable
using high-precision photometry or interferometry. Such observations are required to better understand the phenomenon and to establish ways of mitigating
its impact on BW distances.
Furthermore, only contemporaneous linear and angular radius variations should be
used in BW analyses.
Short-period Cepheids should be observed over up to a few weeks duration to ensure good phase coverage from a single observatory. For long-period
Cepheids, excellent phase coverage must be obtained during individual pulsation
cycles.

Several possible effects explaining the phenomenon can be found 
in the literature. However, a longer observational baseline is required to
further investigate the origin of modulations.

\begin{acknowledgements}
  Many thanks are due to everyone who aided in gathering the present
  dataset and, in particular, to those contributing observations: V.~Bonvin, N.~Cantale,
  B.~Chazelas, P.~Dubath, J.~Hagelberg, D.V.~Martin, F.~Motalebi, N.~Mowlavi,
  L.~Palaversa, S.~Peretti, M.~Tewes, A.~Thoul, and A.~Wyttenbach.
  RIA thanks the anonymous referee for valuable comments that improved
  the quality of the manuscript. Useful discussions with N.~Mowlavi, L.~Eyer,
  X.~Dumusque, S.~Zucker, and B.~Holl, as well as P.I.~Anderson's careful
  reading of the manuscript are acknowledged.
  This research has made use of NASA's ADS Bibliographic Services. 
  RIA acknowledges funding from the Swiss NSF.
\end{acknowledgements}


\hyphenation{Post-Script Sprin-ger}\hyphenation{Post-Script
  Sprin-ger}\hyphenation{Post-Script Sprin-ger}

\Online
\listofobjects

\begin{appendix}
\section{Supporting figures}
\begin{figure*}[]
\centering
\includegraphics{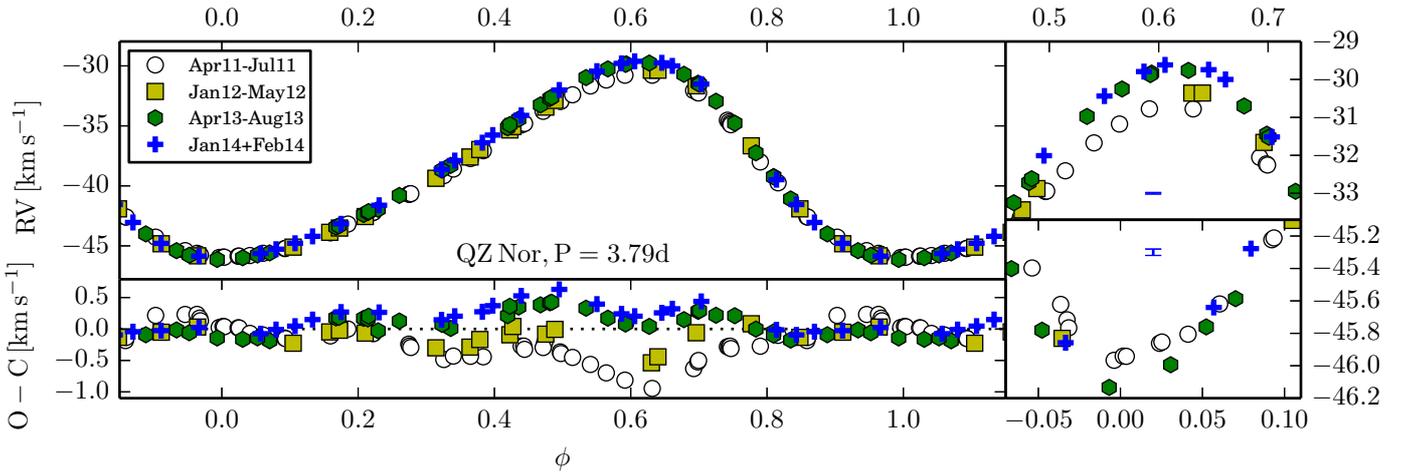}
\caption{New {\it Coralie} RV data for QZ~Nor. The large
panel shows the phased RV curve, distinguishing data from different
epochs by symbol style and color. The
two right-hand panels provide close-ups around maximum (upper) and minimum RV
(lower), and show the median RV uncertainty as a blue errorbar.
The bottom panel shows residuals around an
``average'' Fourier series model fit to the combined data set
\citep[fit procedure described in][]{2013MNRAS.434.2238A}. }
\label{fig:QZNor}
\end{figure*}

\begin{figure*}
\centering
\includegraphics{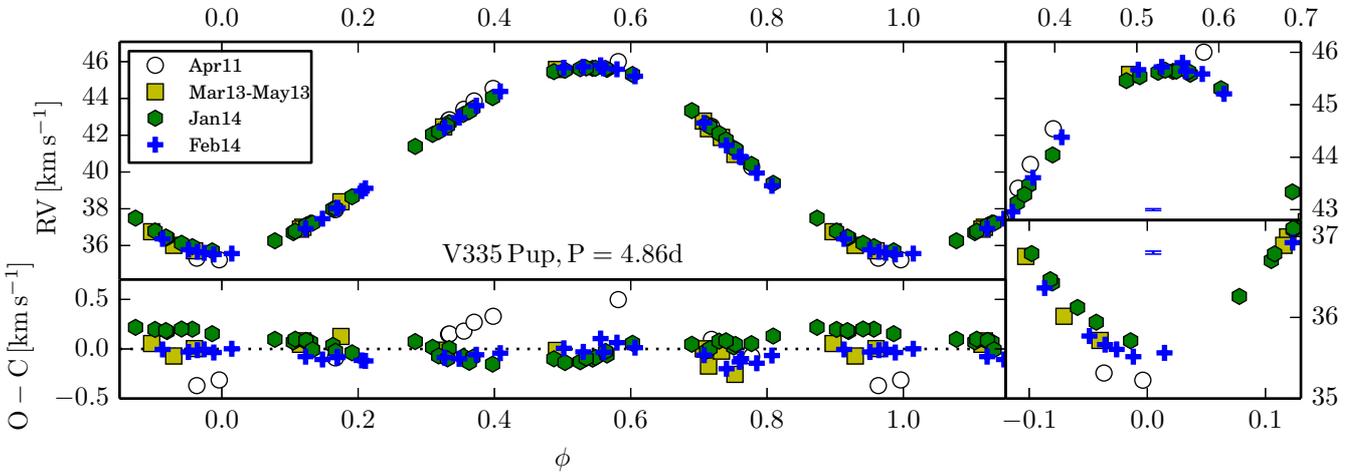}
\caption{Analogous to Fig.\,\ref{fig:QZNor} using RV data for
V335~Puppis.}
\label{fig:V335Pup}
\end{figure*}

\begin{figure*}
\centering
\includegraphics{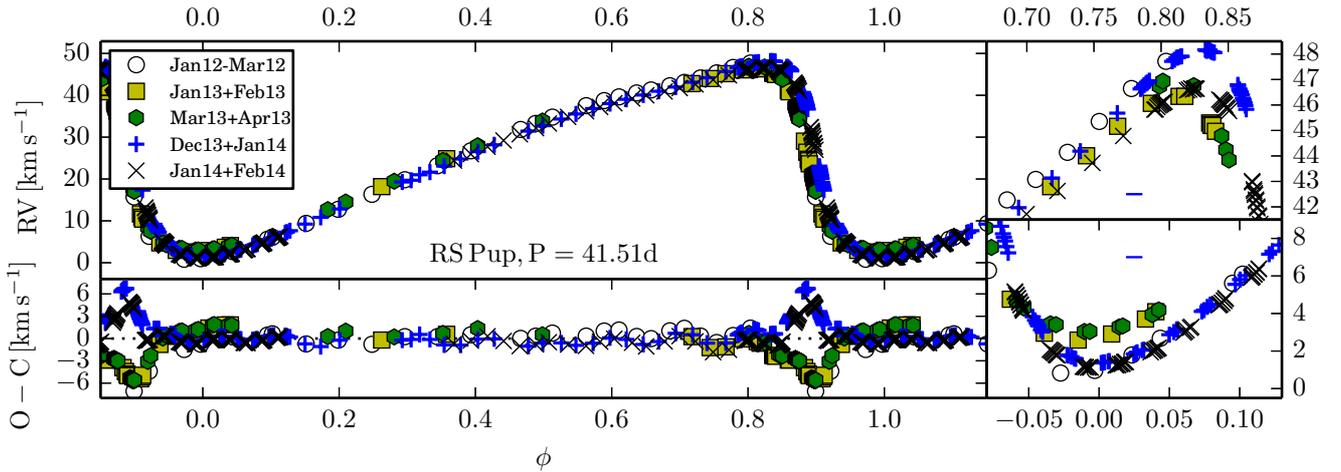}
\caption{Analogous to Fig.\,\ref{fig:QZNor} using RV data for
RS~Puppis. RS~Puppis also exhibits significant random
cycle-to-cycle fluctuations in pulsation period \citep{2009AstL...35..406B}.
As the close-up panels demonstrate, these random period fluctuations occur 
\emph{in addition} to the modulation of the RV amplitude.}
\label{fig:RSPup}
\end{figure*}

This appendix contains Figs. \ref{fig:QZNor}, \ref{fig:V335Pup}, and
\ref{fig:RSPup} that are analogous to Fig.\,\ref{fig:lCar}. These figures 
unambiguously show the modulations discovered. The full dataset employed
to create these figures will be made publicly available through the CDS. 

\begin{figure*}
\centering
\includegraphics{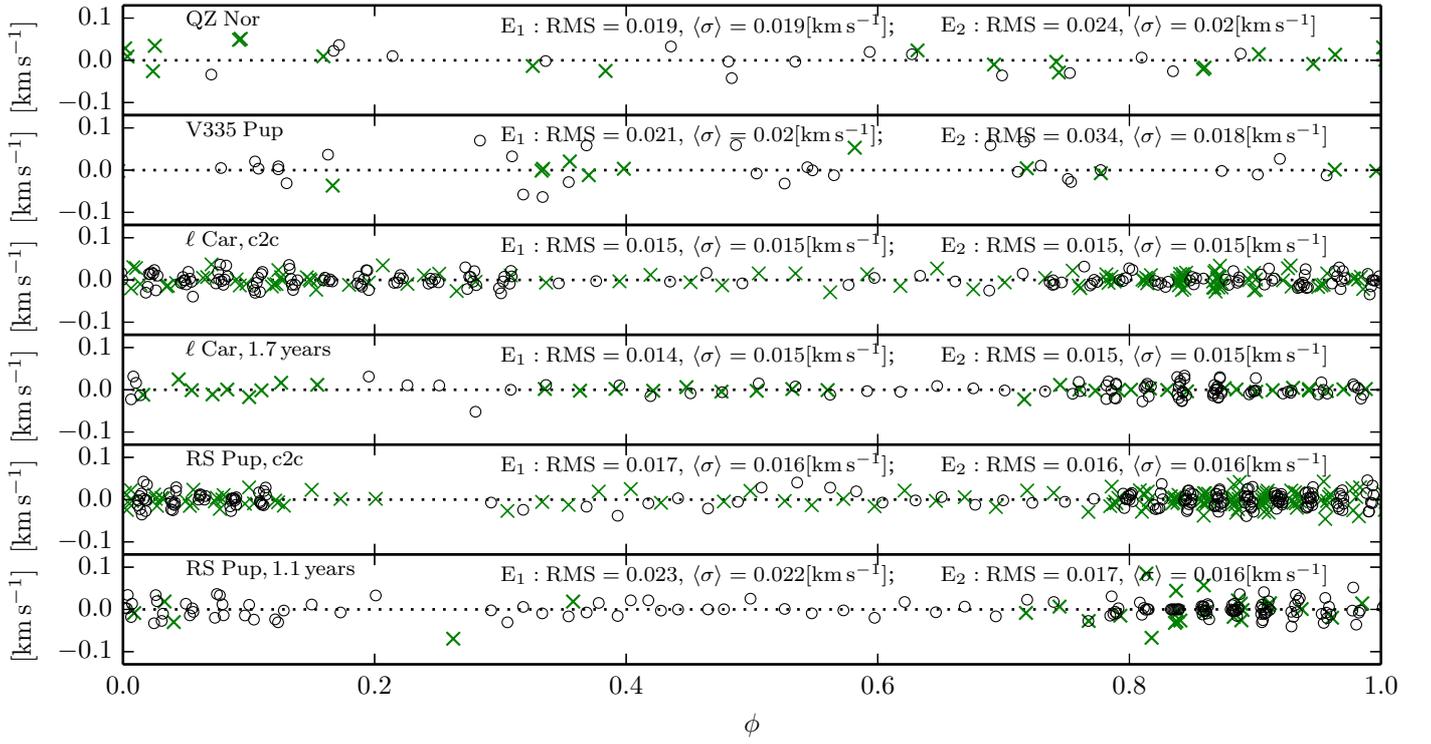}
\caption{Fit residuals from the epochs (E$_1$ and E$_2$) used to
estimate the impact on BW distances for all four Cepheids as listed in
Table\,\ref{tab:obs}. For $\ell$~Car and RS~Pup, both the cycle-to-cycle and
longer timescales are shown. The earlier epoch is represented by green x
markers, the later epoch by black open circles. For each epoch, the RMS around
the fit and median measurement uncertainty, $\langle \sigma \rangle$, are
printed in the corresponding panel, indicating excellent fits.}
\label{fig:residuals}
\end{figure*}

Finally, Fig.\,\ref{fig:residuals} shows the residuals of the models fit to the
per-epoch data used to estimate the impact of modulation as a systematic
uncertainty for Baade-Wesselink distances as listed in Table\,\ref{tab:obs}. The
data were modeled as cubic splines for QZ~Nor, $\ell$~Car, and RS~Pup, and as a
second-order Fourier series for the {\it s}-Cepheid V335~Pup.


\end{appendix}

\end{document}